%% file: main.tex
\documentclass[sigconf]{acmart}

\renewcommand\footnotetextcopyrightpermission[1]{} 
\def\BibTeX{{\rm B\kern-.05em{\sc i\kern-.025em b}\kern-.08emT\kern-.1667em\lower.7ex\hbox{E}\kern-.125emX}}
\usepackage{graphicx}
\usepackage{subcaption}
\usepackage{algorithmic}
\usepackage{algorithm}

\usepackage{amssymb}
\usepackage{pifont}
\newcommand{\cmark}{\ding{51}}%
\newcommand{\xmark}{\ding{55}}%
    
\begin{document}

%
\title{Predicting Drug Responses by Propagating Interactions through Text-Enhanced Drug-Gene Networks}

%



 


\author{Shiyin Wang}
\affiliation{\institution{Massachusetts Institute of Technology}}
\email{shiyinw@mit.edu}


%
\renewcommand{\shortauthors}{Wang, et al.}

%
\begin{abstract}
Personalized drug response has received public awareness in recent years. How to combine gene test result and drug sensitivity records is regarded as essential in the real-world implementation. Research articles are good sources to train machine predicting, inference, reasoning, etc. In this project, we combine the patterns mined from biological research articles and categorical data to construct a drug-gene interaction network. Then we use the cell line experimental records on gene and drug sensitivity to estimate the edge embeddings in the network. Our model provides white-box explainable predictions of drug response based on gene records, which achieve 94.74\% accuracy in binary drug sensitivity prediction task.
\end{abstract}

%
%

\ccsdesc[500]{Information systems}
\ccsdesc[500]{Information systems~Retrieval tasks and goals}
\ccsdesc[100]{Information systems~Information extraction}

\newtheorem{defn}{Definition}
%
\keywords{network science, data mining, drug response}

\maketitle

\input{1_beginning.tex}

\input{2_skeleton.tex}

\input{3_individual.tex}

\input{4_conclusion.tex}

\begin{acks}
Thank Dr. Qi Li for supporting meta-pattern results on PubTator.
\end{acks}

\bibliographystyle{ACM-Reference-Format}
\input{main.bbl}


\end{document}

%% file: 1_beginning.tex
\section{Introduction}

Precision medicine depends on discovering the correlations and casualties between observed data(gene, past clinical records, etc) and unobserved future performance(side effects, drug effects, immune response, etc). Categorical relations have been collected from experiments, while more complex meta-pattern information is mined from scientific publications. To bridge the gap between these two types of data collecting sources, we propose a relation embedding method to expand the representation space of the relationships between genes and drug responses. To the best of my knowledge, this is the first approach to combine text patterns into the existing network. Nodes in the network represent to drug responses(sensitivities, side effects, etc.) and genes.

The model pipeline consists of two parts. First, we construct a relation network of gene and drug response from existing formatted data sources. In the same time, we mine data from scientific publications automatically, retrieving meta-patterns indicating the relations between gene and drug responses. Finally, a novel graph convolutional neural network based method is applied to balance the information collected in the two types of data sources.

\section{Related Works}

Drug discovery plays an essential role in the improvement of human life\cite{drews2000drug} as the pharmacology had become a well-defined and respective scientific discipline. Dating back to the 20th century, researchers across distinct disciplinary areas, including analytical chemistry and biochemistry, demonstrated the values for the drug response analysis. At that time, researchers have already cast interests in the variation of personal drug response. However, because of the limitation of data collection and analysis models, precision medicine did not arouse widespread concern until these decades\cite{mirnezami2012preparing}. Recent years, integrating big data, including clinical data, genetic data, genomic data, intervention history, etc., has received expectations for facilitating characterization of side effects and predicting drug resistance.

Network Science results are applied to the data of gene expression profiles by integrating personalized data to predictive records by similarity analysis, which reveal the hidden correlations of unobserved patterns\cite{Menche:2017ji, Yang:2018gm}. Intuitive speaking, people sharing similar genetic patterns and clinical treatments tend to show similar drug responses. Network-based approaches reveal potential drug-biomarker correlations effectively for some diseases. However, current methods in this aspect are limited in the correlation modeling of the network, and the data accessed is limited in the structured data.

Data Mining communities have been trying to discover, identify, structure, and summary relationships between biological entities through various text mining techniques across open-access biological research publications\cite{sigdel2019cloud, wang2018penner, shang2018automated, wang2018open}. One of the drawbacks of this approach is the accuracy of the results is not high enough to play a deterministic role in the decision process. Instead, they are provided to the human experts as an additional reference to fasten the knowledge discovery process. On the other hand, Recursive Neural Networks has shown its power to deal with text data\cite{lipton2015critical}, demonstrating considerable success in numerous tasks such as image captioning\cite{devlin2015language,you2016image,rennie2017self} and machine translation\cite{cho2014learning,bahdanau2014neural,luong2015effective}. This technique can be applied to convert text meta-patterns into trainable embeddings to represent the relations among entities.

\section{Model}

In this section, we make definitions and formalize the problem. The whole project is consist of two parts: constructing network skeleton and inference edge embedding.

In the first part, we first collected all the entities Definition. \ref{defn:entity} from multiple data sources. Then we mined the categorical relationships and descriptive relationships Definition.\ref{defn:relation}. Patterns are extracted based on the discovered entities and relationships. After that, meta-patterns are extracted from descriptive patterns to be a high-level representation of entities relations. Regarding entities as nodes in the network, we add edges between two nodes if they appear to have either structured or descriptive relationships.

\begin{defn}[Entity]\label{defn:entity}
Entities, denoted by $e$, refer to the names of chemicals, genes, or diseases which can be indexed by a MESH id or Entrez id. For example, SAH is a disease entity with MESH id D013345.
\end{defn}

\begin{defn}[Relation]\label{defn:relation}
Relationships, denoted by $r$, represent the latent interactions between two entities $e_i$, $e_j$. Relations can be either categorical(such as ``blocker", ``binder") or descriptive(such as ``With the increase of $e_i$, there is a significantly drop of $e_j$ expression level").
\end{defn}

\begin{defn}[Pattern]\label{defn:pattern}
The pattern, denoted by $(r, e_i, e_j)$, is a tuple of one relationship and two entities. Patterns are categorical or descriptive concerning the type of the corresponding relationships. Generally, patterns are accessed by directly parsing datasets. Meta-Patterns are the simplified version of descriptive patterns.
\end{defn}

In the second part, we estimate the representations of nodes and edges Definition. \ref{defn:repre}. We model the estimated representation of $v_i$ is calculated by the average neighbor entities to multiple edge representations as Equation. \ref{equ:v}.

\begin{defn}[Edge Representation and Node Representation]\label{defn:repre}
Edge representation, denoted as $R_{e_i, e_j}\in \mathcal M$, is the quantified version of relationship defined on a function family. In this project, because of the computation limitation, we set $\mathcal M=\mathcal R_{k\times k}$ to be a matrix of size $k\times k$. Then nodes are represented as a vector $v_i\in \mathcal R^k$, bias $b_i$, and scale $\lambda_i$. The estimated value of $v_i$ is given by:
\begin{equation}\label{equ:v}
    \hat{v_i} = \frac{\lambda_i}{|N(e_i)|} \sum_{e_j \in N(e_i)}R_{e_i, e_j} v_j + \lambda_ib_i
\end{equation}
\end{defn}

Having built the network skeleton, we then need to learn the representation $R_{e_i, e_j}$, $\lambda_i$, $b_i$, $v_i$ of edges and nodes. 

In the datasets, researchers record TPM (transcripts per million) to quantify genes and use intensity to measure drug sensitivity. To convert representations into a numeric format, we deploy a fully connected neural network with one hidden layer of size 2 and sigmoid activation.

\begin{equation}
    u_i = S(v_i)
\end{equation}

We define the loss function as:
\begin{equation}
    \mathcal L_{gene} = \frac{1}{N_{gene}}\sum_{i\in Gene} (\hat{u_i} - u_i)^2
\end{equation}
\begin{equation}
    \mathcal L_{chem} = \frac{1}{N_{gene}}\sum_{i\in Chemical} (\hat{u_i} - u_i)^2
\end{equation}
\begin{equation}\label{equ:loss}
    \mathcal L = \mathcal L_{gene} + \beta \mathcal L_{chem}
\end{equation}

\begin{algorithm}
\begin{algorithmic}
\caption{Representation Learning}
\label{code}
\FOR{Epoch $1, 2, \dots, N$}
\FOR{Iteration $1, 2, \ldots, T$}
\STATE Calculate $\hat{v_i}=G(v_i, b_i, \lambda_i, R_{e_i, e_j})$ by Equation. \ref{equ:v}
\STATE Calculate $\mathcal L$ by Equation. \ref{equ:loss}
\STATE Update $v_i\gets v_i - \alpha \frac{\partial \mathcal L}{\partial v_i}$
\STATE Update $b_i\gets b_i - \alpha \frac{\partial \mathcal L}{\partial b_i}$
\STATE Update $\lambda_i\gets v_i - \alpha \frac{\partial \mathcal L}{\partial \lambda_i}$
\ENDFOR
\STATE Calculate $\hat{v_i}=G(v_i, b_i, \lambda_i, R_{e_i, e_j})$ by Equation. \ref{equ:v}
\STATE Calculate $\mathcal L$ by Equation. \ref{equ:loss}
\STATE Update $R_{e_i, e_j}\gets v_i - \alpha \frac{\partial \mathcal L}{\partial R_{e_i, e_j}}$
\ENDFOR
\end{algorithmic}
\end{algorithm}

%% file: 2_skeleton.tex
\section{Implementation}
\subsection{Collecting Structured Data}

Drug-Gene interactions are well studied in the past biological researches, such as CancerCommons, CF:Biomarkers, CGI, NCI, DrugBank, PharmGKB, DoCM, etc. SIDER\footnote{\url{sideeffects.embl.de}} provides adverse drug reactions (ADRs) representing drug responses, which contains 1430 drugs, 5880 ADRs and 140,064 drug-ADR pairs. The Cancer Therapeutics Response Portal (CTRP)\footnote{\url{portals.broadinstitute.org/ctrp/}} links genetic, lineage, and other cellular features of cancer cell lines to small-molecule sensitivity with the goal of accelerating the discovery of patient-matched cancer therapeutics. In this paper, we use the drug-gene interaction database (DGIdb)\cite{pmid26656090, pmid23993102, seashore2015harnessing}, which contains drug-gene interactions and gene druggability information from 30 different sources. Figure \ref{fig:count_plot} shows the frequency of drug-gene interaction types in DGIdb, showing a long-tailed distribution. The inhibitor occurs 9982 times in DGIdb, which is 40.79\% of the whole interaction records. The second most frequent interaction is agonist, which occurs 5333 times and accounts for 21.79\%. There are 30 types of relationships. We assume the founded structured relations are correct and precise.

\subsection{Mining Descriptive Relationship}

PubMed abstracts are a good source to obtain descriptive patterns. We investigated a subset of annotated PubMed data by PubTator\cite{wei2013pubtator}. We selected all the sentences that contain two annotated chemicals or genes. The total size of the file is approximate 2 gigabyte, which is too complex to take the original long sentences into the model. The first reason is that many patterns may occur a few times if we use sentences to model the relations directly. Secondly, our computation power does not have enough memory to process large graphs. Therefore, we further process our descriptive patterns to short representations, which we call meta-patterns.

\begin{figure*}[t!]
    \centering
    \caption{An example of pattern and meta-pattern extraction process from a sentence in a PubMed paper.}
    \includegraphics[width=\linewidth]{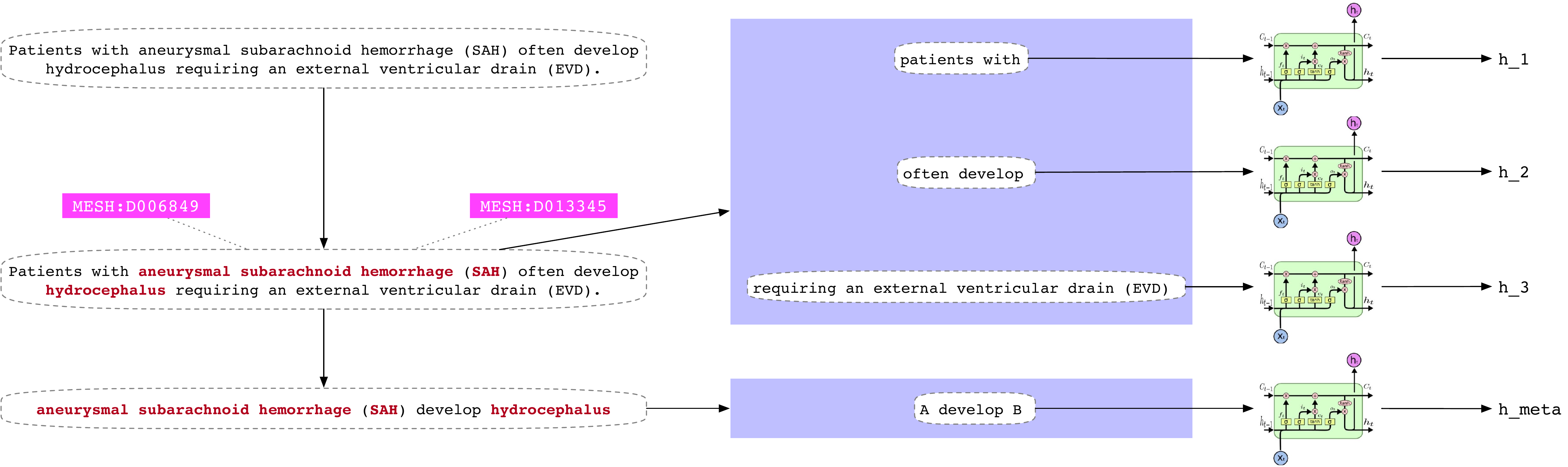}
    \label{fig:metapatternexp}
\end{figure*}

\subsection{Extract Meta-Patterns}

For decades, numerous biological researchers describe their findings on interactions between chemicals, drugs, genes, etc., by natural languages. The adequate biological research literature provide a good source to extract the information using information retrieval methods\cite{Shang:2017wl, :2017uo, Shang:2016wf}. We collected 3614796 abstracts from PubMed and used the PubTator dataset to discover all the named entities in chemicals, diseases, and genes. Meta pattern\cite{Wang:2018cl, :2017uo} is defined as a frequent, informative, and precise subsequence pattern in certain context. The first step of meta-pattern discovery is context-aware segmentation, which estimates the contextual boundaries of meta-patterns. We use the method developed by Wang\cite{Wang:2018cl}, which extracted relationships from a subset of the abstracts on the PubMed\cite{Davis:2017fx} with the semi-supervise from CTD database\footnote{\url{http://ctdbase.org}}. For example, the meta-patterns in this sentence from a PubMed abstract are shown in the Figure. \ref{fig:metapatternexp}


\subsection{Linking Categorical Relations and Descriptive Relations}

One of the challenges in this project is the entity linking across multiple datasets, which refer to the alignment of entity mentions to its entity id. This process is very tedious because of the messy index format. For example, chemicals(drugs) have CID, PMID, Chembl, Cas Number, etc. We use the entity name as a validation criterion. Though the same entity may have different names in different data sources, those names are similar with high probability. Therefore, we can measure the correctness of our entity linking algorithm by comparing merged names. The summary of all the data sources used in this project are listed in Table. \ref{tab:dataset} \ref{tab:dataset2}.

\begin{table}
  \caption{Datasets summary}
  \label{tab:dataset}
  \begin{tabular}{ccccl}
    \toprule
    Dataset & \# Gene & \# Drug & \# Relations & Format\\
    \hline
    \hline
    DGIdb &36815 & 9370 & 42727 & Categorical\\
    PubTator &36815 & 9370 & 42727 & Descriptive\\
    PubMed\footnote{Meta-patterns extracted from a subset of the abstracts on the PubMed.} & 2575 & 6199 & 10530 & MetaPattern \\
    RNA-seq\cite{Barretina:2012fp} & 58035 & - & - & CL Records\\
    GDSC\cite{Yang:2013kx}  & - & 16 & - & CL Records\\
    \hline
    Skeleton &335 & 587 & 3321 & - \\
  \bottomrule
\end{tabular}
\end{table}

\begin{table}[H]
  \caption{Datasets}
  \label{tab:dataset2}
  \begin{tabular}{p{1cm}p{6cm}}
    \toprule
    Dataset & URL\\
    \hline
    \hline
    DGIdb & http://www.dgidb.org\\
    PubTator &https://www.ncbi.nlm.nih.gov/CBBresearch/Lu/ Demo/PubTator \\
    PubMed & https://www.ncbi.nlm.nih.gov/pubmed\\
    RNA-seq\cite{Barretina:2012fp} & https://www.ebi.ac.uk/gxa/experiments/E-MTAB-2770/Results\\
    GDSC\cite{Yang:2013kx}  & https://www.cancerrxgene.org \\
    \hline
  \bottomrule
\end{tabular}
\end{table}

\begin{figure*} 
  \label{fig:pairs} 
  \caption{The visualization of all mined interactions among genes and drugs. The label of the genes are their Entrez ID. The label of chemicals are their names. (\# drugs=587, \# genes=335, \# interactions=3321)}
  \begin{subfigure}[b]{0.45\linewidth}
    \centering
    \includegraphics[width=\linewidth]{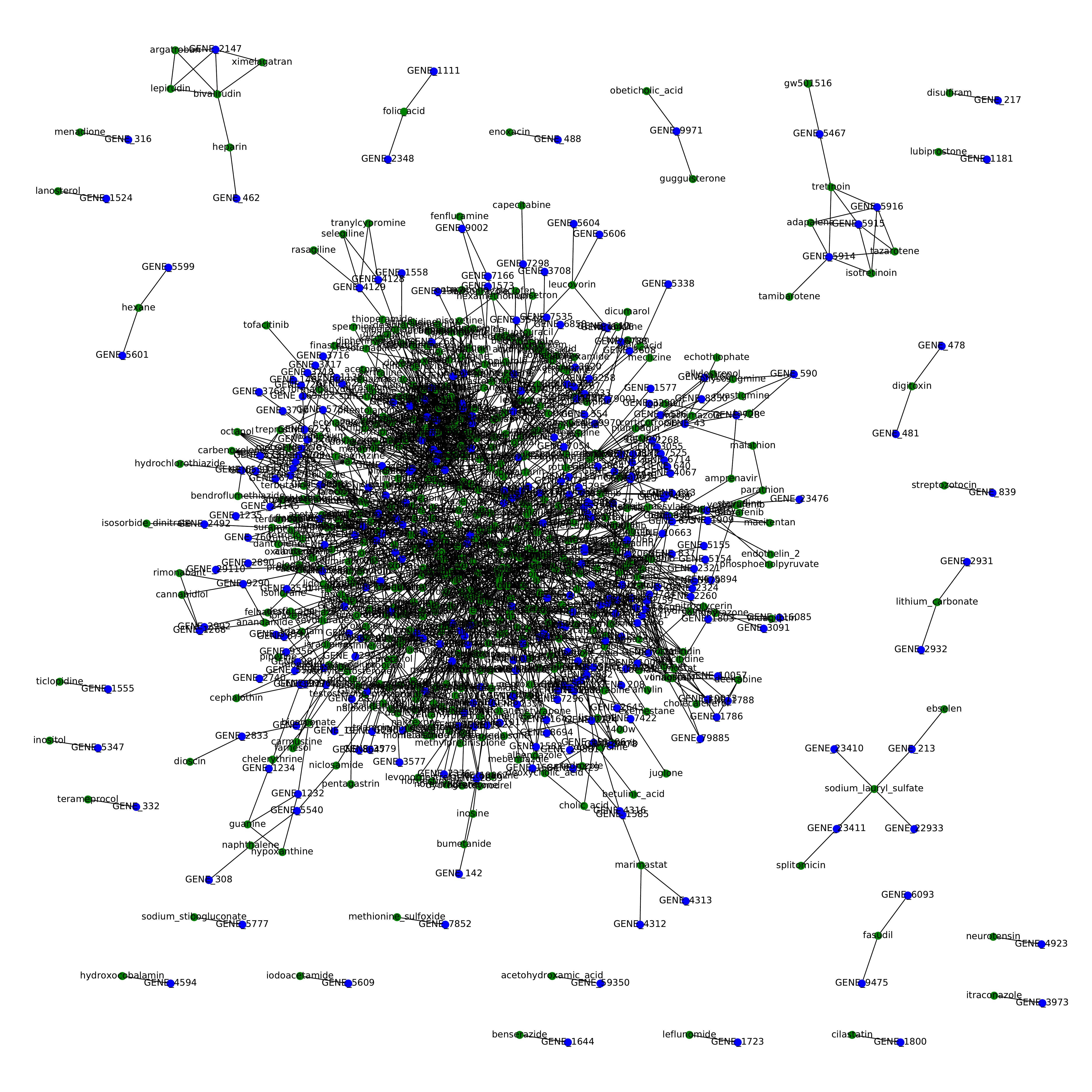} 
    \caption{Display entity names}
  \end{subfigure}
  \begin{subfigure}[b]{0.45\linewidth}
    \centering
    \includegraphics[width=\linewidth]{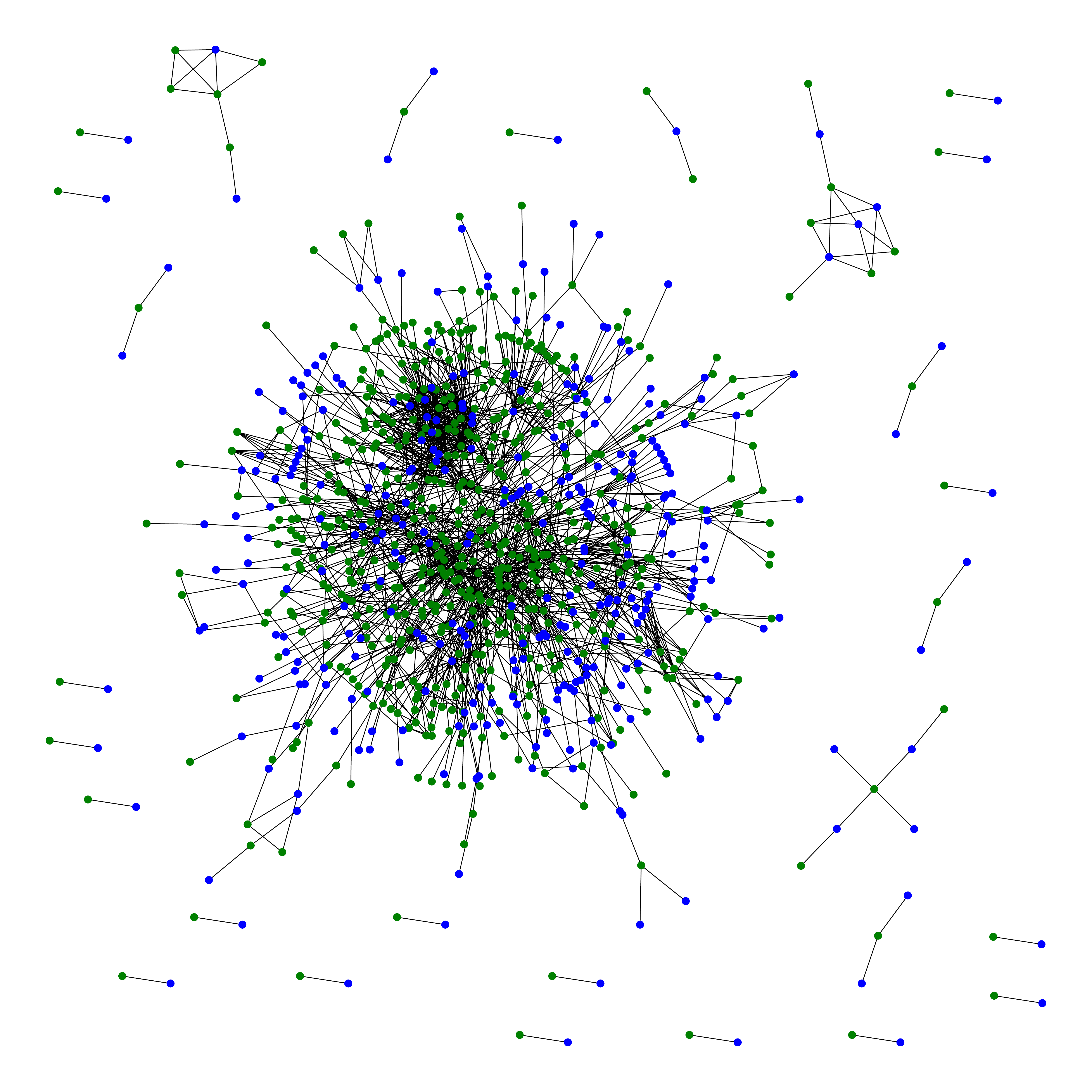} 
    \caption{Not display entity names}
  \end{subfigure} 
\end{figure*}

%% file: 3_individual.tex

%% file: 4_conclusion.tex
\subsection{Inference Individual Drug Responses}

To determine the parameters in the model, we acquire cell line experiment records of 58035 gene expression on 934 human cancer cell lines \cite{Barretina:2012fp} and 3723759 drug sensitivity test records on 1065 cell lines\cite{Yang:2013kx}. We assume the same cell lines perform similarly in these two data sources. Then we can align these two data sources by cell line names. The data summary is shown in Table. \ref{tab:dataset}.

\begin{figure*}
    \centering
    \label{fig:4pic}
    \caption{28 entities are contained in cell line records, which are colored with red. We extracted them and their neighbors within distance k. Then we pruned unimportant unobserved entities.}\label{fig:prune}
    \begin{subfigure}[b]{0.22\textwidth}
        \includegraphics[width=\textwidth]{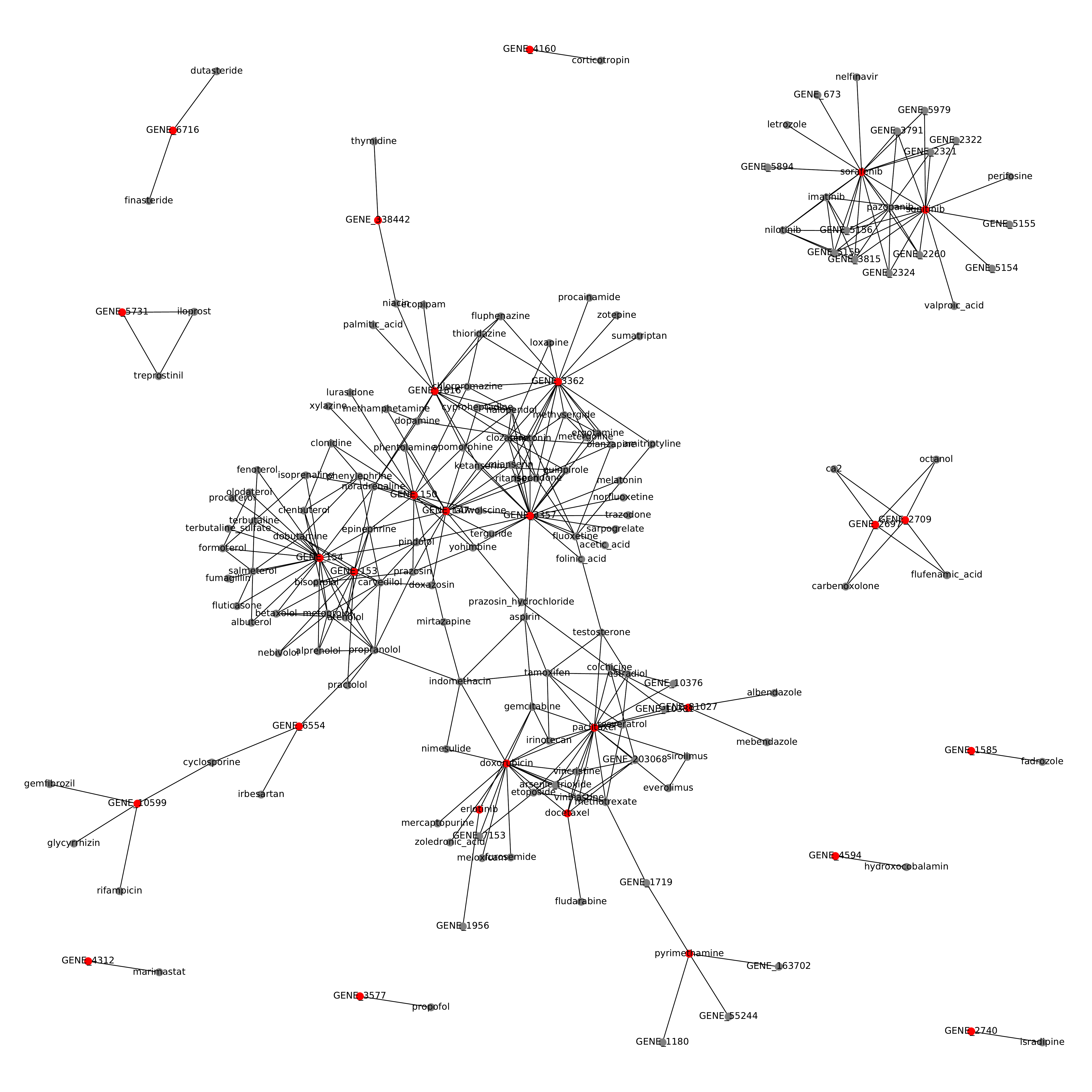}
        \caption{k=1}
        \label{fig:small1}
    \end{subfigure}
    ~ 
    \begin{subfigure}[b]{0.22\textwidth}
        \includegraphics[width=\textwidth]{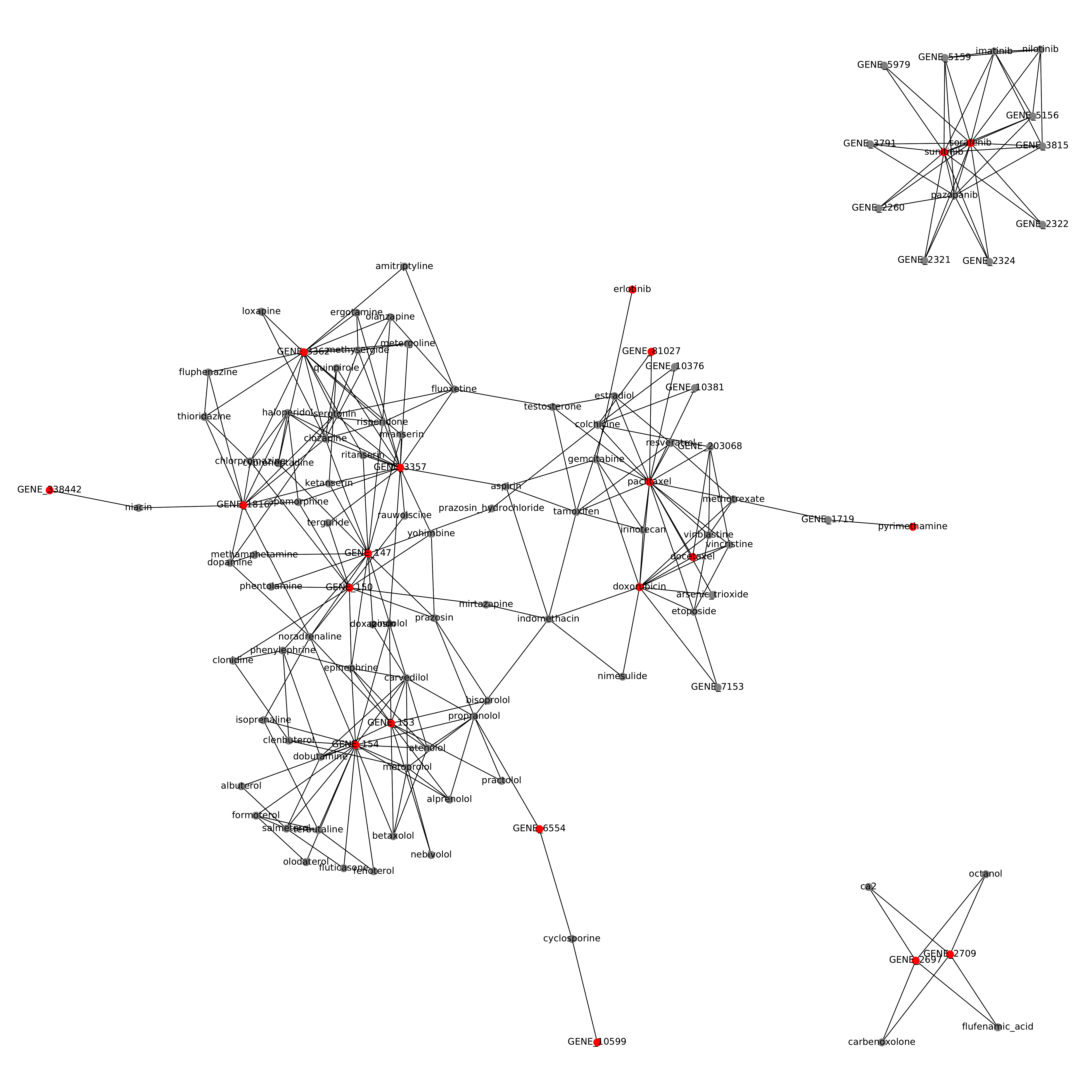}
        \caption{k=1, pruned}
        \label{fig:small1p}
    \end{subfigure}
    ~ 
    \begin{subfigure}[b]{0.22\textwidth}
        \includegraphics[width=\textwidth]{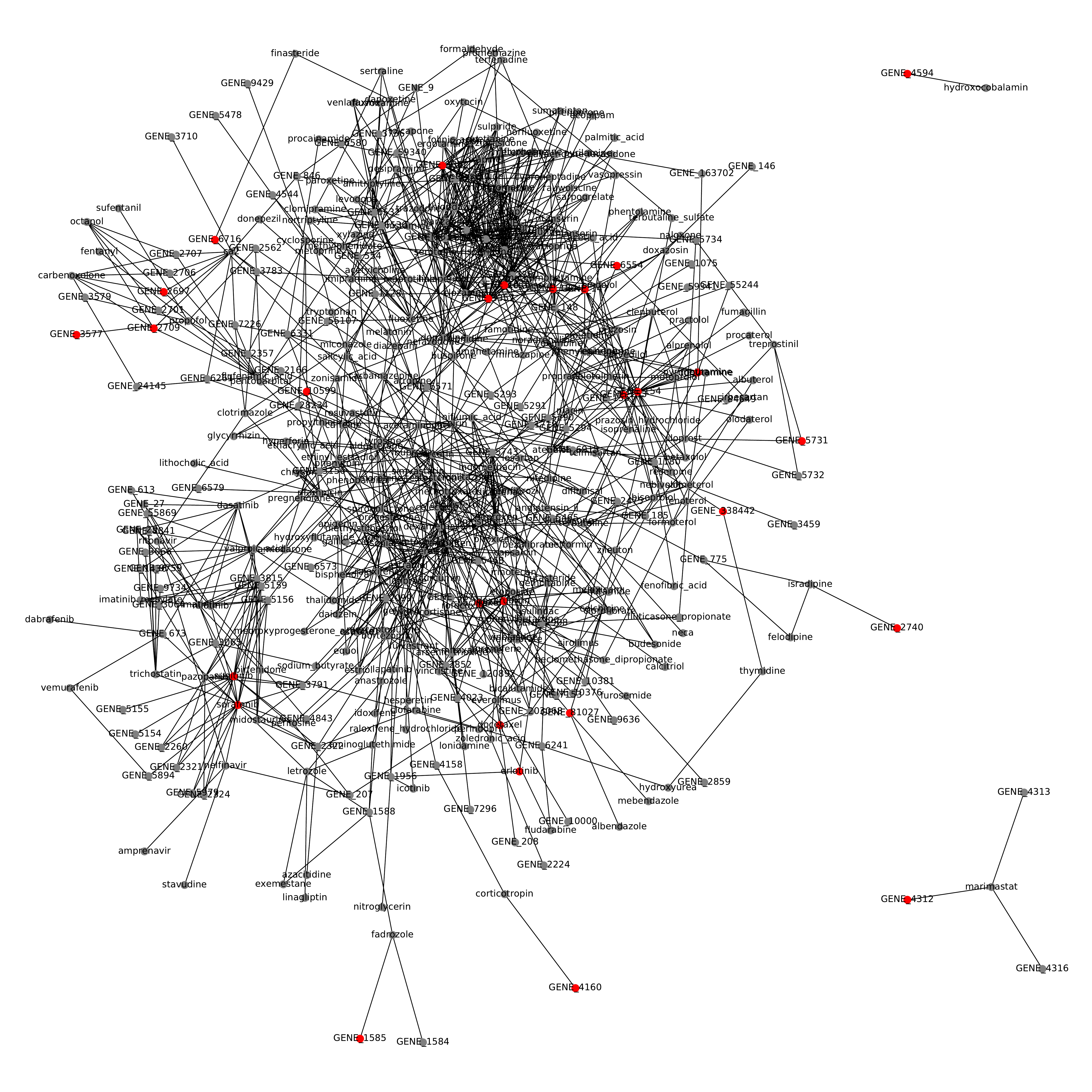}
        \caption{k=2}
        \label{fig:small2}
    \end{subfigure}
    \begin{subfigure}[b]{0.22\textwidth}
        \includegraphics[width=\textwidth]{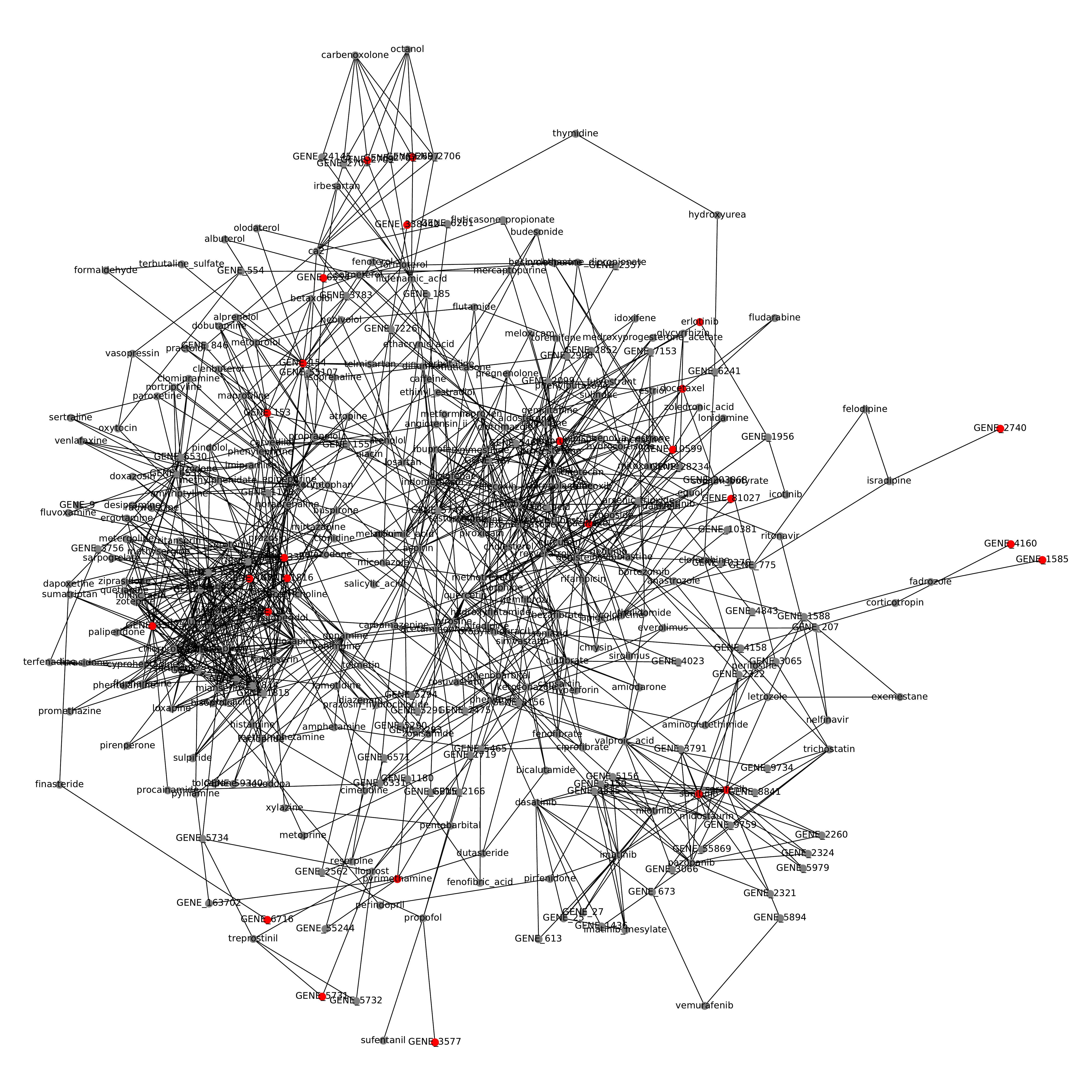}
        \caption{k=2, pruned}
        \label{fig:small2p}
    \end{subfigure}
\end{figure*}

Because we have to use experimental data to train the representations, we need to shrink the graph, deleting useless nodes. We pick the nodes from observable entities(red) with distance 1 or 2 to the observable entities. The results are shown on Figure. \ref{fig:small1}, \ref{fig:small1p}, \ref{fig:small2}, \ref{fig:small2p}.

We set the embedding dimension to be 4. The learning rate of gene optimizer is 0.1 so that it can converge quickly in \ref{code}. The learning rate of chemical is 0.001.

We have done experiments on the network of Figure. \ref{fig:small1p}. The binary classification accuracy is 91.15\%. Because network data is not structured, it is hard to draw a ROC or some other training visualization.

\section{Evaluation}

We test our model on the extracted 539 drug-gene records linked by cell-line. The data contain test records of 7 drugs and 21 genes, and about 42.12\% cells are missing. The inadequacy of data limits the performance of complex models. We compare with Logistic Regression and Support Vector Machine. We use sklearn package for Logistic Regression, with the L1 penalty and max interation 10000. We use radial basis function kernel in SVM model with auto gamma. Our method outperforms these two baselines.

Our model can be explainable because it is based on the representation of categorical and descriptive relations. Human experts can guide the model by changing the network skeleton. Therefore, it can potentially avoid the risk of black-box decision and provide an interpretation of its decisions.

\begin{center}
\begin{table}[!hbt]
  \caption{Model Performance}
  \label{sample-table}
  \centering
  \begin{tabular}{c|cccc}
    \toprule
    Method & Logistic Reg. & SVM & Ours\\
    \hline
Accuracy & 93.75\% & 90.14\% & 94.74\%\\
Explainable & \xmark & \xmark & \cmark \\
    \bottomrule
  \end{tabular}
\end{table}
\end{center}

\section{Future Work}

\paragraph{Collect More Records} The experiment records are not adequate to test our model in scale. Patient profiles will be a good data source for our model. But due to the privacy restrictions, we do not have them right now. The application of our model on clinical data can also help to provide precise and interpretable patient profiles.

\paragraph{Attention Mechanism} Because of the limitation of gene-drug records, we do not apply sophisticated design in this model. Intuitively, the drug responses will be better modeled if we allow the weight of different relations to change according to their current value.

\section{Conclusion}

Research articles are excellent sources for machines to predict, infer, reasoning, etc. By aggregating multiple data sources, we study and analyze the association of drug and gene by constructing networks with categorical relations and descriptive relations. The resulted network skeleton presents the relationships between entities. We model the relations as a kernel function between entities. After that, we use cell line experiment records to estimate the parameters of our model, which achieves 94.74\% accuracy on predicting the drug sensitivity for cell lines.